\documentclass[12pt]{article}

\input{tcilatex}

\begin{document}

\begin{center}
{\Large Hamiltonian Noether theorem for gauge systems}

\smallskip

{\Large and two time physics}

\textbf{\vspace{0.6cm}}

V. M. Villanueva$^{\ddag ,}$\footnote{%
E-mail: vvillanu@ifm.umich.mx}, J. A. Nieto$^{\dag ,}$\footnote{%
E-mail: \texttt{nieto@uas.uasnet.mx}}, L. Ruiz$^{\dag }$ and J. Silvas$%
^{\dag }$

\vspace{0.6cm}

$^{\ddag }$\emph{Instituto de F\'{\i}sica y Matem\'{a}ticas}\\[0pt]
\emph{Universidad Michoacana de San Nicol\'{a}s de Hidalgo}\\[0pt]
\emph{P.O. Box 2-82, Morelia, Michoac\'{a}n, M\'{e}xico}\\[0pt]
\vspace{0.5cm}

$^{\dag }$\emph{Escuela de Ciencias Fisico-Matem\'{a}ticas}\\[0pt]
\emph{Universidad Aut\'{o}noma de Sinaloa}\\[0pt]
\emph{Culiac\'{a}n, Sinaloa, M\'{e}xico}\\[0pt]
\vspace{0.5cm}

\bigskip\ 

\textbf{Abstract}
\end{center}

The Noether theorem for Hamiltonian constrained systems is revisited. In
particular, our review presents a novel method to show that the gauge
transformations are generated by the conserved quantities associated with
the first class constraints. We apply our results to the relativistic point
particle, to the {Friedberg \textit{et al}. model and, with special
emphasis, to two time physics.}

\bigskip\ 

\bigskip\ 

\bigskip\ 

\bigskip\ 

\bigskip\ 


Pacs numbers: 04.60.-m, 04.65.+e, 11.15.-q, 11.30.Ly

April, 2005

\newpage

\noindent \textbf{1.- Introduction}

\smallskip\ 

The main motivation of these notes is to revisit the Hamiltonian approach of
the Noether theorem [1] in the case of singular systems. Our formalism is
focused entirely on the Hamiltonian sector and we do not attempt to describe
the corresponding Lagrangian sector in the sense of Gr\`{a}cia and Pons
[2]-[4] construction.

We work out in a fundamental constrained Hamiltonian formalism [5]-[8],
which is characterized by a first order functional action defined on the
phase space variables $q^{i}$ and $p_{j},$ with $i,j=1,\dots ,n$. Consider
the first class canonical Hamiltonian $H(q,p;t)$ and all the first class
constraints of the system $\phi _{\alpha }(q,p;t)$ with their respective
Lagrange multipliers $\lambda ^{\alpha }(t)$. The first order action is

\begin{equation}
S[q,p;\lambda ^{\alpha }]=\int_{t_{i}}^{t_{f}}dt[\dot{q}^{i}p_{i}-H(q,p;t)-%
\lambda ^{\alpha }(t)\phi _{\alpha }(q,p;t)],  \tag{1}
\end{equation}%
where $\dot{q}^{i}=\frac{d}{dt}q^{i}$ and $\alpha =1,2,\dots ,r.$ Here, the
Lagrange multipliers $\lambda ^{\alpha }(t)$ are not regarded as dynamical
degrees of freedom, but only as auxiliary variables which parametrize the
gauge degrees of freedom of the system. In fact, it is not difficult to
prove that if one considers the Lagrange multipliers as dynamical variables,
then their associated canonical momenta $\pi _{\alpha }$ are first class
constraints which only lead to arbitrary shifts in the Lagrange multipliers,
in total agreement with their auxiliary character and do not act on the
phase space variables [6].

We shall assume that in this fundamental Hamiltonian formalism, all second
class constraints, if any, have been solved and implemented in the dynamics
of the system in such a way that only first class constraints are involved
in the gauge invariance as well as on the dynamical evolution of the system
[6]-[7].

Our main task is to obtain Noether's first and second theorems [6] for gauge
systems in the Hamiltonian sector (see Refs. [9]-[13]). In particular, when
the second theorem is applied, we conclude that the conserved quantities are
precisely the first class constraints. It is important to mention that in
the case of regular systems (free of constraints) all the well known results
are obtained.

We first apply our formalism to two examples: The relativistic point
particle, and the Friedberg \textit{et al.} model [14] which has been
studied and solved for the case of Grivov ambiguities in [15]-[16]. We show
that in these cases the conserved quantities are precisely the first class
constraints.

It turns out that two time physics (see [17] and Refs. therein) offers
another interesting example for applying our formalism. The main reason is
that in two time physics the variables $q^{i}$ and $p^{j}$ are unified in
just one object $x_{a}^{i}$, with $a=1,2$, where $x_{1}^{i}\equiv q^{i}$%
\textbf{\ }and $x_{2}^{i}\equiv p^{i}$, and consequently in the
corresponding action the hidden symmetry $Sp(2,R)$ or $SL(2,R)$ becomes
manifest (see Refs. [18]-[22]). Thus, we show that our formalism shed some
new light on this hidden symmetry.

This work is organized as follows. In Section 2, we discuss the Hamiltonian
Noether theorem for gauge systems. In Sections 3 and 4, we describe the
Noether's first and second theorems respectively. In Section 5, we apply our
procedure to two examples: the relativistic scalar particle and the helix
model of Friedberg \textit{et al.}. In Section 6, we discuss two time
physics from the point of view of our formalism. Finally, in Section 7 we
make some final remarks.

\bigskip\ 

\noindent \textbf{2.-} \textbf{Hamiltonian Noether theorem for gauge systems}

\smallskip\ 

Let us first rewrite the action (1) in the form

\begin{equation}
S[q,p;\lambda ^{\alpha }]=\int_{t_{i}}^{t_{f}}dt\left[ \dot{q}^{i}p_{i}-H_{T}%
\right] ,  \tag{2}
\end{equation}%
where

\begin{equation}
H_{T}=H(q,p;t)+\lambda ^{\alpha }(t)\phi _{\alpha }(q,p;t)  \tag{3}
\end{equation}%
denotes de total Hamiltonian. Our aim is to see the consequences of applying
to the action (2) the total variations:

\begin{equation}
\delta t=t^{\prime }(t)-t,  \tag{4}
\end{equation}%
\begin{equation}
\delta _{\star }q^{i}=q^{\prime i}(t^{\prime })-q^{i}(t)=\delta q^{i}+\dot{q}%
^{i}\delta t,  \tag{5}
\end{equation}%
\begin{equation}
\delta _{\star }p_{i}=p_{i}^{\prime }(t^{\prime })-p_{i}(t)=\delta p_{i}+%
\dot{p}_{i}\delta t  \tag{6}
\end{equation}%
and%
\begin{equation}
\delta _{\star }\lambda ^{\alpha }=\lambda ^{\prime \alpha }(t^{\prime
})-\lambda ^{\alpha }(t)=\delta \lambda ^{\alpha }+\dot{\lambda}^{\alpha
}\delta t,  \tag{7}
\end{equation}%
where $\delta q^{i}=q^{\prime i}(t)-q^{i}(t)$ and similar expressions hold
for $\delta p_{i}$ and $\delta \lambda ^{\alpha }$. Observe that the
expression (5) for $\delta _{\star }q^{i}$ implies

\begin{equation}
\delta _{\star }\dot{q}^{i}=\delta \dot{q}^{i}+\ddot{q}^{i}\delta t.  \tag{8}
\end{equation}%
It is important to remark that $\delta \dot{q}^{i}=\frac{d}{dt}\delta q^{i}$
but $\delta _{\star }\dot{q}^{i}\neq \frac{d}{dt}\delta _{\star }q^{i}$.

Invariance of the action (2) under total variations means that

\begin{equation}
\delta _{\star }S=\int_{t_{i}}^{t_{f}}dt\frac{d}{dt}\delta _{\star }\Lambda
(q,p),  \tag{9}
\end{equation}%
where $\Lambda (q,p)$ is an arbitrary function. Thus, using transformations
(4)-(7) we obtain

\begin{equation}
\begin{array}{c}
\delta _{\star }S=\int_{t_{i}}^{t_{f}}dt\delta _{\star }[\dot{q}%
^{i}p_{i}-H_{T}]+\int_{t_{i}}^{t_{f}}dt\frac{d\delta t}{dt}[\dot{q}%
^{i}p_{i}-H_{T}] \\ 
\\ 
=\int_{t_{i}}^{t_{f}}dt\frac{d}{dt}\delta _{\star }\Lambda (q,p).%
\end{array}
\tag{10}
\end{equation}%
It is not difficult to show that the expression (10) leads to

\begin{equation}
\int_{t_{i}}^{t_{f}}dt\left\{ \frac{d}{dt}Q+\dot{q}^{i}\delta _{\star }p_{i}-%
\dot{p}_{i}\delta _{\star }q^{i}+\delta t\dot{H}_{T}-\delta _{\star
}H_{T}\right\} =0.  \tag{11}
\end{equation}%
Here, the variable $Q=Q(q,p;t)$ is defined as

\begin{equation}
Q=\delta _{\star }q^{i}p_{i}-\delta tH_{T}-\delta _{\star }\Lambda . 
\tag{12}
\end{equation}%
In virtue of the definitions of the total variations (4)-(7) we find that
the relation (11) can also be written as

\begin{equation}
\int_{t_{i}}^{t_{f}}dt\left\{ \frac{d}{dt}Q+\dot{q}^{i}\delta p_{i}-\dot{p}%
_{i}\delta q^{i}-\delta H_{T}\right\} =0.  \tag{13}
\end{equation}%
Let us write (13) in the form

\begin{equation}
A+B=0,  \tag{14}
\end{equation}%
where

\begin{equation}
A=\int_{t_{i}}^{t_{f}}dt\frac{d}{dt}Q  \tag{15}
\end{equation}%
and%
\begin{equation}
B=\int_{t_{i}}^{t_{f}}dt\left\{ \dot{q}^{i}\delta p_{i}-\dot{p}_{i}\delta
q^{i}-\delta H_{T}\right\} .  \tag{16}
\end{equation}%
The expression (14), or (13), offers three different possibilities, namely

(1) If $A=0$ then (13) implies that $B=0$.

(2) If $B=0$ then (13) implies that $A=0.$

(3) If neither $A$ nor $B$ are zero then (13) establishes that $A+B=0.$

\noindent The first two cases are well known, but the third seems to have
passed unnoticed. In order to clarify these observations let us briefly
discuss each one of these three cases. In the first case, we assume that the
quantity $Q$ satisfies the expression

\begin{equation}
Q\mid _{t_{i}}^{t_{f}}=0,  \tag{17}
\end{equation}%
which is equivalent to say that $\int_{t_{i}}^{t_{f}}dt\frac{d}{dt}Q=0$ or $%
A=0$. Thus, for arbitrary variations, $\delta q^{i},$ $\delta p_{i}$ and $%
\delta \lambda ^{\alpha }$, (13) yields

\begin{equation}
\int_{t_{i}}^{t_{f}}dt\big\{\big(\dot{q}^{i}-\frac{\partial }{\partial p_{i}}%
H_{T}\big)\delta p_{i}+\big(-\dot{p}_{i}-\frac{\partial }{\partial q^{i}}%
H_{T}\big)\delta q^{i}-\delta \lambda ^{\alpha }\phi _{\alpha }\big \}=0, 
\tag{18}
\end{equation}%
and therefore we get the equations of motion:

\begin{equation}
\dot{q}^{i}=\frac{\partial }{\partial p_{i}}H_{T}=\big\{q^{i},H_{T}\big \}, 
\tag{19}
\end{equation}

\begin{equation}
\dot{p}_{i}=-\frac{\partial }{\partial q^{i}}H_{T}=\big\{p_{i},H_{T}\big \} 
\tag{20}
\end{equation}%
and

\begin{equation}
\phi _{\alpha }=0.  \tag{21}
\end{equation}%
Here, the symbol $\big\{f,g\big \}$, for any functions $f$\ and $g$\ of the
canonical variables $q^{i}$ and $p_{i}$, stands for the usual Poisson
bracket, that is

\begin{equation}
\big\{f,g\big \}=\frac{\partial f}{\partial q^{i}}\frac{\partial g}{\partial
p_{i}}-\frac{\partial g}{\partial q^{i}}\frac{\partial f}{\partial p_{i}}. 
\tag{22}
\end{equation}

In the second case, we assume that the dynamical system satisfies equations
of motion (19)-(21). This means that (18) follows which means that $B=0.$
Therefore from (13) we see that

\begin{equation}
\int dt\frac{d}{dt}Q=0.  \tag{23}
\end{equation}%
Since by hypothesis the interval $t_{f}-t_{i}$ is arbitrary, from (23) we
have $\frac{d}{dt}Q=0$ and therefore we find that $Q$ is a conserved
quantity.

The last possibility arises if we assume that neither (18) nor (23) hold,
that is, we assume that $A$ and $B$ are different from zero. We shall show
that in this case the expression (13) implies that $Q$ is the generator of
canonical transformations. For this purpose let us first compute $\frac{d}{dt%
}Q$. Since $Q=Q(q,p;t),$ we have

\begin{equation}
\frac{d}{dt}Q=\frac{\partial Q}{\partial q^{i}}\dot{q}^{i}+\frac{\partial Q}{%
\partial p_{i}}\dot{p}_{i}+\frac{\partial Q}{\partial t}.  \tag{24}
\end{equation}%
Thus, for an undefined interval $t_{f}-t_{i},$ (13) becomes

\begin{equation}
\int dt\bigg \{\frac{\partial Q}{\partial q^{i}}\dot{q}^{i}+\frac{\partial Q%
}{\partial p_{i}}\dot{p}_{i}+\frac{\partial Q}{\partial t}+\dot{q}^{i}\delta
p_{i}-\dot{p}_{i}\delta q^{i}-\delta H_{T}\bigg\}=0,  \tag{25}
\end{equation}%
which can be rewritten as

\begin{equation}
\int dt\bigg \{\left( \frac{\partial Q}{\partial q^{i}}+\delta p_{i}\right) 
\dot{q}^{i}+\left( \frac{\partial Q}{\partial p_{i}}-\delta q^{i}\right) 
\dot{p}_{i}+\frac{\partial Q}{\partial t}-\delta H_{T}\bigg\}=0,  \tag{26}
\end{equation}%
or

\begin{equation}
\int \bigg \{\left( \frac{\partial Q}{\partial q^{i}}+\delta p_{i}\right)
dq^{i}+\left( \frac{\partial Q}{\partial p_{i}}-\delta q^{i}\right)
dp_{i}+\left( \frac{\partial Q}{\partial t}-\delta H_{T}\right) dt\bigg\}=0.
\tag{27}
\end{equation}

If we now define the quantity

\begin{equation}
\omega =\left( \frac{\partial Q}{\partial q^{i}}+\delta p_{i}\right)
dq^{i}+\left( \frac{\partial Q}{\partial p_{i}}-\delta q^{i}\right)
dp_{i}+\left( \frac{\partial Q}{\partial t}-\delta H_{T}\right) dt,  \tag{28}
\end{equation}%
we observe that (27) gives

\begin{equation}
\int \omega =0.  \tag{29}
\end{equation}%
From (28) we observe that $\omega $ may admit an interpretation of 1-form.
Thus, under usual assumptions (29) implies that $\omega $ is an exact form
which means that

\begin{equation}
\omega =df,  \tag{30}
\end{equation}%
where $f$ is an arbitrary zero-form.

We shall assume that $f=f(q,p)$. From (28) and (27) we see that

\begin{equation}
\frac{\partial Q}{\partial t}-\delta H_{T}=0.  \tag{31}
\end{equation}%
Thus, considering (31) we discover that the expressions (27) and (30) yield

\begin{equation}
\left( \frac{\partial Q^{\prime }}{\partial q^{i}}+\delta p_{i}\right)
dq^{i}+\left( \frac{\partial Q^{\prime }}{\partial p_{i}}-\delta
q^{i}\right) dp_{i}=0,  \tag{32}
\end{equation}%
where

\begin{equation}
Q^{\prime }=Q+f.  \tag{33}
\end{equation}%
Since, $dq^{i}$ and $dp_{i}$ are $1-$form bases we find that (32) implies

\begin{equation}
\delta q^{i}=\frac{\partial Q^{\prime }}{\partial p_{i}}=\big\{%
q^{i},Q^{\prime }\big \}  \tag{34}
\end{equation}%
and%
\begin{equation}
\delta p_{i}=-\frac{\partial Q^{\prime }}{\partial q^{i}}=\big\{%
p_{i},Q^{\prime }\big \}.  \tag{35}
\end{equation}%
Thus, we have shown that up to an arbitrary function $f$ the quantity $Q,$
which is a conserved quantity when the equations of motion are satisfied, is
the generator of canonical transformations.

In order to clarify the meaning of expression (31), we investigate the
consequences of invariances under gauge transformations, \textit{i.e.,} we
consider the particular case%
\begin{equation}
Q^{\prime }=\xi ^{\alpha }(t)\phi _{\alpha }(q,p;t),  \tag{36}
\end{equation}%
where the quantities $\xi ^{\alpha }(t)$ are infinitesimal parameters
associated with the first class constraints $\phi _{\alpha }(q,p;t)$.
Moreover; since we are dealing (by assumption) with only first class
constraints $\phi _{\alpha }(q,p;t)$, we can write (see Refs. [6] and [23])

\begin{equation}
\big\{H,\phi _{\alpha }\big \}=V_{\alpha }^{\beta }\phi _{\beta }  \tag{37}
\end{equation}%
and%
\begin{equation}
\big\{\phi _{\alpha },\phi _{\beta }\big \}=C_{\alpha \beta }^{\gamma }\phi
_{\gamma },  \tag{38}
\end{equation}%
where $V_{\alpha }^{\beta }$ and $C_{\alpha \beta }^{\gamma }$ are structure
"constants". Then, (31), (34), (35) and (36) lead to%
\begin{equation}
\delta \lambda ^{\alpha }\phi _{\alpha }=\left( \dot{\xi}^{\alpha }-\xi
^{\beta }V_{\beta }^{\alpha }-\xi ^{\beta }\lambda ^{\gamma }C_{\beta \gamma
}^{\alpha }\right) \phi _{\alpha }.  \tag{39}
\end{equation}%
Considering that the first class constraints $\phi _{\alpha }(q,p;t)$ are
independent functions we get that the expression (39) implies%
\begin{equation}
\delta \lambda ^{\alpha }=\dot{\xi}^{\alpha }-\xi ^{\beta }V_{\beta
}^{\alpha }-\xi ^{\beta }\lambda ^{\gamma }C_{\beta \gamma }^{\alpha }, 
\tag{40}
\end{equation}%
which is the usual result for the transformations of the Lagrange
multipliers $\lambda ^{\alpha }$ under gauge transformations generated by
the first class constraints $\phi _{\alpha }(q,p;t)$ (see Refs. [6], [7] and
[23]).

\bigskip\ 

\noindent \textbf{3.-Noether's first theorem}

\smallskip\ 

We now consider the consequences of previous discussion for the particular
case of transformations that define a simply connected continuous group. In
other words, we are interested in studying the so called first Noether
theorem which refers to the invariance of the action (2) under global
transformations. Of course, these transformations are not associated with a
gauge physical system because this arises when one assumes local
transformations. Thus, we consider the transformations%
\begin{equation}
\begin{array}{c}
\delta t=\xi ^{\alpha }\chi _{\alpha }(t), \\ 
\\ 
\delta q^{i}=\xi ^{\alpha }\varphi _{\alpha }^{i}(q,p;t), \\ 
\\ 
\delta p_{i}=\xi ^{\alpha }\psi _{i\alpha }(q,p;t), \\ 
\\ 
\delta \Lambda =\xi ^{\alpha }\Lambda _{\alpha }(q,p;t),%
\end{array}
\tag{41}
\end{equation}%
with the abbreviations%
\begin{equation}
\begin{array}{c}
\varphi _{\alpha }^{i}=\{q^{i},\phi _{\alpha }\}, \\ 
\\ 
\psi _{i\alpha }=\{p_{i},\phi _{\alpha }\}, \\ 
\\ 
\Lambda _{\alpha }=\{\Lambda ,\phi _{\alpha }\}.%
\end{array}
\tag{42}
\end{equation}%
Here, $\xi ^{\alpha },$ with $\alpha =1,2,\dots ,r,$ is an infinitesimal
constant parameter spanning infinitesimal group transformations, with $r$ as
the dimension of such a group.

Using (41) we find that the expression (13) gives%
\begin{equation}
\begin{array}{c}
\int_{t_{i}}^{t_{f}}dt\xi ^{\alpha }\bigg \{\frac{d}{dt}\left[ \varphi
_{\alpha }^{i}p_{i}+\dot{q}^{i}p_{i}\chi _{\alpha }-\chi _{\alpha
}H_{T}-\Lambda _{\alpha }-\dot{\Lambda}\chi _{\alpha }\right] \\ 
\\ 
+\left( \dot{q}^{i}-\frac{\partial }{\partial p_{i}}H_{T}\right) \psi
_{i\alpha }+\left( -\dot{p}_{i}-\frac{\partial }{\partial q^{i}}H_{T}\right)
\varphi _{\alpha }^{i} \\ 
\\ 
-\left( V_{\alpha }^{\beta }-\lambda ^{\gamma }C_{\alpha \gamma }^{\beta
}\right) \phi _{\beta }\bigg\}=0.%
\end{array}
\tag{43}
\end{equation}%
Thus, according to our discussion of the previous section we find that the
relation (43) determines that, up to an arbitrary function, the quantity $%
Q=\xi ^{\alpha }Q_{\alpha },$ where%
\begin{equation}
Q_{\alpha }=\varphi _{\alpha }^{i}p_{i}+\dot{q}^{i}p_{i}\chi _{\alpha }-\chi
_{\alpha }H_{T}-\Lambda _{\alpha }-\dot{\Lambda}\chi _{\alpha }  \tag{44}
\end{equation}%
are the $r$-conserved Noether charges that in turn, generate the
transformations (41).

\pagebreak

\noindent \textbf{4.-} \textbf{Noether's second theorem}

\smallskip\ 

As a second application, we now consider the case in which the parameters of
transformation $\xi ^{\alpha }$ are functions of the time $t$. In addition,
we assume that the corresponding gauge transformations are generated by the
first class constraints $\phi _{\alpha }$ [6] (see Ref. [9] and [10] for
details),%
\begin{equation}
\begin{array}{c}
\delta t=\xi ^{\alpha }(t)\chi _{\alpha }(t), \\ 
\\ 
\delta q^{i}=\xi ^{\alpha }(t)\varphi _{\alpha }^{i}(q,p;t), \\ 
\\ 
\delta p_{i}=\xi ^{\alpha }(t)\psi _{i\alpha }(q,p;t), \\ 
\\ 
\delta \Lambda =\xi ^{\alpha }(t)\Lambda _{\alpha }(q,p;t), \\ 
\\ 
\delta \lambda ^{\alpha }=\dot{\xi}^{\alpha }(t)-\xi ^{\beta }(t)V_{\beta
}^{\alpha }-\xi ^{\beta }(t)\lambda ^{\gamma }C_{\beta \gamma }^{\alpha },%
\end{array}
\tag{45}
\end{equation}%
where we used the definitions (42). It is important to mention that, in
these expressions, we have not considered other possible derivatives of $\xi
^{\alpha }(t)$. Nevertheless, the generalization to such cases seems to be
straightforward.

The substitution of relations (45) into expression (13) yields to%
\begin{equation}
\begin{array}{c}
\int_{t_{i}}^{t_{f}}dt\bigg \{\frac{d}{dt}\left[ \xi ^{\alpha }(t)Q_{\alpha
}(q,p;t)\right] \\ 
\\ 
+\xi ^{\alpha }(t)\bigg[\left( \dot{q}^{i}-\frac{\partial }{\partial p_{i}}%
\big(H+\lambda ^{\rho }\phi _{\rho }\big)\right) \psi _{i\alpha } \\ 
\\ 
+\left( -\dot{p}_{i}-\frac{\partial }{\partial q^{i}}\big(H+\lambda ^{\rho
}\phi _{\rho }\big)\right) \varphi _{a}^{i}\bigg] \\ 
\\ 
-\dot{\xi}^{\alpha }(t)\phi _{\alpha }-\xi ^{\alpha }V_{\alpha }^{\rho }\phi
_{\rho }-\xi ^{\alpha }\lambda ^{\gamma }C_{\alpha \gamma }^{\rho }\phi
_{\rho }\bigg\}=0,%
\end{array}
\tag{46}
\end{equation}%
with%
\begin{equation}
Q_{\alpha }=\varphi _{a}^{i}p_{i}+\dot{q}^{i}p_{i}\chi _{\alpha }-\chi
_{\alpha }\big(H+\lambda ^{\rho }\phi _{\rho }\big)-\Lambda _{\alpha }. 
\tag{47}
\end{equation}%
Thus, according to the discussion in Section 2 the $Q_{\alpha }$ given in
(47) can be associated with the conserved charges or the generator of the
gauge transformations (45) depending on whether the corresponding equations
of motion of the gauge physical system are satisfied or not respectively.

\bigskip\ 

\noindent \textbf{5.- Relativistic point particle and Friedberg \textit{et
al.} model}

\smallskip\ 

\noindent $\bullet $ \textbf{Example 1}{: The relativistic scalar point
particle.}

The general covariant Hamiltonian formulation of the free relativistic point
particle of mass $m_{0}$ is provided by the phase space $(x^{\mu },p_{\nu
}), $ with Poisson brackets%
\begin{equation}
\big\{x^{\mu },p_{\nu }\big\}=\delta _{\nu }^{\mu },  \tag{48}
\end{equation}%
and the first class constraint%
\begin{equation}
\phi =\frac{1}{2}\big(p^{2}+m_{0}^{2}\big).  \tag{49}
\end{equation}%
The total Hamiltonian $H_{T}$ is given by 
\begin{equation}
H_{T}=\frac{\lambda }{2}\big(p^{2}+m_{0}^{2}\big),  \tag{50}
\end{equation}%
where $\lambda $ is the Lagrange multiplier associated with\textbf{\ }the
first class constraint (49).

The corresponding fundamental first order action is%
\begin{equation}
S[x,p;\lambda ]=\int_{\tau _{i}}^{\tau _{f}}d\tau \left[ \dot{x}^{\mu
}p_{\mu }-\frac{\lambda }{2}\big(p^{2}+m_{0}^{2}\big)\right] .  \tag{51}
\end{equation}%
This action is invariant (up to a surface term) under the transformation
generated by the first class constraint%
\begin{equation}
\begin{array}{c}
\delta x_{\mu }=\big\{x_{\mu },\xi \phi \big\}={\xi }p_{\mu }, \\ 
\\ 
\delta p^{\mu }=\big\{p^{\mu },\xi \phi \big\}=0, \\ 
\\ 
\delta \lambda =\dot{\xi}.%
\end{array}
\tag{52}
\end{equation}%
In fact, using (52) we get%
\begin{equation}
\delta S=\int_{\tau _{i}}^{\tau _{f}}d\tau \frac{d}{d\tau }\left[ {\frac{\xi 
}{2}}\big(p^{2}-m_{0}^{2}\big)\right] ,  \tag{53}
\end{equation}%
which leads to the surface term

\begin{equation}
\delta \Lambda =\xi \Lambda _{1}=\frac{{\xi }}{2}\big(p^{2}-m_{0}^{2}\big). 
\tag{54}
\end{equation}%
Thus, according to (41) we see that from (52) we can conclude that $\chi
_{1}=0,$ $\varphi _{1}^{\mu }=p^{\mu },$ $\psi _{\mu 1}=0.$ Using these
results and $\Lambda _{1}$ given in (54) we find that the expression (44)
implies that our conserved quantity is%
\begin{equation}
Q=\xi (\varphi _{1}^{\mu }p_{\mu }-\Lambda _{1})={\frac{\xi }{2}}\big(%
p^{2}+m_{0}^{2}\big),  \tag{55}
\end{equation}%
as expected.

\bigskip\ 

\noindent $\bullet $ \textbf{Example 2}{: The Friedberg \textit{et al.}
model.}

The helix model of Friedberg \textit{et al.} [14] (see also Refs. [15] and
[16]) can be described in terms of the fundamental Hamiltonian first order
action:

\begin{equation}
\begin{array}{c}
S=\int_{t_{i}}^{t_{f}}dt\bigg[\dot{x}p_{x}+\dot{y}p_{y}+\dot{z}p_{z}-{\frac{1%
}{2}}\big[p_{x}^{2}+p_{y}^{2}+p_{z}^{2}\big] \\ 
\\ 
-U(x,y)-\lambda \big(p_{z}+g\big(xp_{y}-yp_{x}\big)\big)\bigg],%
\end{array}
\tag{56}
\end{equation}%
where $(x,y,z)$ and $(p_{x},p_{y},p_{z})$ stand for three dimensional
coordinates and canonical momenta respectively. Where $U(x,y)=U\big(%
x^{2}+y^{2}\big)$, and $\lambda $ is a Lagrange multiplier associated with
the first class constraint $\phi =p_{z}+g\big(xp_{y}-yp_{x}\big)$, where $g$
denotes a coupling constant.

This action is invariant under the infinitesimal gauge transformations%
\begin{equation}
\begin{array}{c}
\delta x=-\alpha y, \\ 
\\ 
\delta y=+\alpha x, \\ 
\\ 
\delta z=+{\frac{1}{g}}\alpha%
\end{array}
\tag{57}
\end{equation}%
and%
\begin{equation}
\begin{array}{c}
\delta p_{x}=-\alpha p_{y}, \\ 
\\ 
\delta p_{y}=+\alpha p_{x}, \\ 
\\ 
\delta p_{z}=0.%
\end{array}
\tag{58}
\end{equation}%
Furthermore, we have

\begin{equation}
\delta \lambda =+{\frac{1}{g}}\dot{\alpha}.  \tag{59}
\end{equation}%
Thus, from (57), (58) and (59) we find the following identifications: $\xi
^{1}(t)={\frac{1}{g}}\alpha (t),$ $\chi _{1}=0,$ $\varphi _{1}^{1}=-gy,$ $%
\varphi _{1}^{2}=+gx\ $and$\ \varphi _{1}^{3}=1,$ as well as $\psi
_{11}=-gp_{y},$ $\psi _{21}=+gp_{x},$ and $\psi _{31}=0.$ Observe that in
this case, the variation of the action is exactly zero and there is no need
for the surface term as in the previous example. With the above ingredients,
by direct substitution in (44), we get

\begin{equation}
\begin{array}{l}
Q_{1}=\varphi _{1}^{i}p_{i}+0, \\ 
\\ 
=-gyp_{x}+gxp_{y}+1p_{z}, \\ 
\\ 
=p_{z}+g\big (xp_{y}-yp_{x}\big),%
\end{array}
\tag{60}
\end{equation}%
which is precisely the first class constraint of the physical system whose
motion is governed by the action (56).

\bigskip\ 

\noindent \textbf{6.- Two time physics}

\smallskip\ 

Two time physics is described by the action [17] (see also Refs. [18]-[22])

\begin{equation}
S=\int_{\tau _{i}}^{\tau _{f}}d\tau \left( \frac{1}{2}\varepsilon ^{ab}\dot{x%
}_{a}^{\mu }x_{b}^{\nu }\eta _{\mu \nu }-H(x_{a}^{\mu })\right) ,  \tag{61}
\end{equation}%
where $\eta _{\mu \nu }$ is a flat metric whose signature will be determined
below. Up to a total derivative this action is equivalent to the first order
action

\begin{equation}
S=\int_{\tau _{i}}^{\tau _{f}}d\tau \left( \dot{x}^{\mu }p_{\mu
}-H(x,p)\right) ,  \tag{62}
\end{equation}%
where

\begin{equation}
\begin{array}{c}
x^{\mu }=x_{1}^{\mu }, \\ 
\\ 
p^{\mu }=x_{2}^{\mu }.%
\end{array}
\tag{63}
\end{equation}

For a relativistic point particle one chooses $H$ as $H_{T}=\lambda (p^{\mu
}p_{\mu }+m_{0}^{2})$ (see example 1 in Section 5) or

\begin{equation}
H_{T}=\lambda (p^{\mu }p_{\mu }),  \tag{64}
\end{equation}%
in the massless case. Observing that the first term in the action (61) has
the manifest $Sp(2,R)$ (or $SL(2,R)$) invariance, we find that these choices
for $H$ spoil such a symmetry for the entire action (61).

It turns out that the simplest possible choice for $H$ which maintains the
symmetry $Sp(2,R)$ is

\begin{equation}
H=\frac{1}{2}\lambda ^{ab}x_{a}^{\mu }x_{b}^{\nu }\eta _{\mu \nu },  \tag{65}
\end{equation}%
where $\lambda ^{ab}=\lambda ^{ba}$ is a Lagrange multiplier. One may also
think in the "massive" case

\begin{equation}
H=\frac{1}{2}\lambda ^{ab}(x_{a}^{\mu }x_{b}^{\nu }\eta _{\mu \nu
}+m_{ab}^{2}),  \tag{66}
\end{equation}%
with $m_{11}^{2}=-R^{2}$, $m_{22}^{2}=m_{0}^{2}$ and $m_{12}^{2}=0$, but one
can see that the mass term $m_{ab}^{2}$ breaks the $Sp(2,R)$-symmetry$.$
Considering (65) the action (61) becomes

\begin{equation}
S=\int_{\tau _{i}}^{\tau _{f}}d\tau \left( \frac{1}{2}\varepsilon ^{ab}\dot{x%
}_{a}^{\mu }x_{b}^{\nu }\eta _{\mu \nu }-\frac{1}{2}\lambda ^{ab}x_{a}^{\mu
}x_{b}^{\nu }\eta _{\mu \nu }\right) .  \tag{67}
\end{equation}%
Arbitrary variations of $\lambda ^{ab}$ in (67) lead to the constraint

\begin{equation}
\Omega _{ab}=x_{a}^{\mu }x_{b}^{\nu }\eta _{\mu \nu }=0,  \tag{68}
\end{equation}%
which turns out to be first class.

In terms of the notation (63) we find that the expression (68) gives (see
Ref. [24])

\begin{equation}
x^{\mu }x_{\mu }=0,  \tag{69}
\end{equation}

\begin{equation}
x^{\mu }p_{\mu }=0  \tag{70}
\end{equation}%
and

\begin{equation}
p^{\mu }p_{\mu }=0.  \tag{71}
\end{equation}%
While in the "massive" case (66) leads to

\begin{equation}
x^{\mu }x_{\mu }-R^{2}=0,  \tag{72}
\end{equation}

\begin{equation}
x^{\mu }p_{\mu }=0  \tag{73}
\end{equation}%
and

\begin{equation}
p^{\mu }p_{\mu }+m_{0}^{2}=0.  \tag{74}
\end{equation}

The key point in two time physics comes from the observation that if $\eta
_{\mu \nu }$ corresponds to just one time, that is, if $\eta _{\mu \nu }$
has the signature $\eta _{\mu \nu }=diag(-1,1,...,1)$ then from (69)-(71) it
follows that $p^{\mu }$ is parallel to $x^{\mu }$ and therefore the angular
momentum

\begin{equation}
L^{\mu \nu }=x^{\mu }p^{\nu }-x^{\nu }p^{\mu }  \tag{75}
\end{equation}%
associated with the Lorentz symmetry of (67) should vanish, which is, of
course, an unlikely result. Thus, if we impose the condition $L^{\mu \nu
}\neq 0$ and the constraints (69)-(71) we find that the signature of $\eta
_{\mu \nu }$ should be at least of the form $\eta _{\mu \nu
}=diag(-1,-1,1,...,1).$ In other words, only with two times the constraints
(69)-(71) are consistent with the requirement $L^{\mu \nu }\neq 0$ (see
Refs. [25]-[27]). In principle we can assume that the number of times is
greater than 2, but then one does not have enough constraints to eliminate
all the possible ghosts.

With these observations at hand, we shall now proceed to generalize the
action (2) in the form

\begin{equation}
S[x_{a}^{\mu };\lambda ]=\int_{\tau _{i}}^{\tau _{f}}d\tau \bigg[\frac{1}{2}%
\varepsilon ^{ab}\dot{x}_{a}^{\mu }x_{b}^{\nu }\eta _{\mu \nu }-H_{T}\bigg].
\tag{76}
\end{equation}%
Here

\begin{equation}
H_{T}=H(x_{a}^{\mu };\tau )-\lambda ^{bc}(\tau )\phi _{bc}(x_{a}^{\mu };\tau
).  \tag{77}
\end{equation}%
We are assuming that $\phi _{bc}=\phi _{cb}$ denotes a generalization of the
first class constraint $\Omega _{ab}$ (see expression (68)).

Consider the transformations

\begin{equation}
\begin{array}{c}
\delta \tau =\tau ^{\prime }(\tau )-\tau , \\ 
\\ 
\delta _{\star }x_{a}^{\mu }=x_{a}^{\prime \mu }(\tau ^{\prime })-x_{a}^{\mu
}(\tau )=\delta x_{a}^{\mu }+\dot{x}_{a}^{\mu }\delta \tau , \\ 
\\ 
\delta _{\star }\lambda _{ab}=\lambda _{ab}^{\prime }(\tau ^{\prime
})-\lambda _{ab}(\tau )=\delta \lambda _{ab}+\dot{\lambda}_{ab}\delta \tau ,%
\end{array}
\tag{78}
\end{equation}

where $\delta x_{a}^{\mu }=x_{a}^{\prime \mu }(\tau )-x_{a}^{\mu }(\tau )$
and a similar expression for $\delta \lambda _{ab}$ holds. The expression
for $\delta _{\star }x_{a}^{\mu }$ implies

\begin{equation}
\delta _{\star }\dot{x}_{a}^{\mu }=\delta \dot{x}_{a}^{\mu }+\ddot{x}%
_{a}^{\mu }\delta \tau .  \tag{79}
\end{equation}

We find that invariance of the action (76) under the transformations (78)
gives

\begin{equation}
\begin{array}{c}
\delta _{\star }S=\int_{\tau _{i}}^{\tau _{f}}d\tau \delta _{\star }\left[ 
\frac{1}{2}\varepsilon ^{ab}\dot{x}_{a}^{\mu }x_{b}^{\nu }\eta _{\mu \nu
}-H_{T}\right] +\int_{\tau _{i}}^{\tau _{f}}d\tau {\frac{d\delta \tau }{%
d\tau }}\left[ \frac{1}{2}\varepsilon ^{ab}\dot{x}_{a}^{\mu }x_{b}^{\nu
}\eta _{\mu \nu }-H_{T}\right] , \\ 
\\ 
=\int_{\tau _{i}}^{\tau _{f}}d\tau {\frac{d}{d\tau }}\delta _{\star }\Lambda
(x_{a}^{\mu };\tau ),%
\end{array}
\tag{80}
\end{equation}%
where $\Lambda (x_{a}^{\mu };\tau )$ is an arbitrary function. It is not
difficult to show that this variation of the action $S$ can be reduced to
the form:

\begin{equation}
\begin{array}{c}
\int_{\tau _{i}}^{\tau _{f}}d\tau \big\{\frac{d}{d\tau }\left[ \frac{1}{2}%
\varepsilon ^{ab}\delta _{\star }x_{a}^{\mu }x_{b}^{\nu }\eta _{\mu \nu
}-\delta \tau H_{T}-\delta _{\star }\Lambda \right] +\varepsilon ^{ab}\dot{x}%
_{a}^{\mu }\delta x_{b}^{\nu }\eta _{\mu \nu } \\ 
\\ 
+\delta \tau \dot{H}_{T}-\delta _{\star }H_{T}\big\}=0.%
\end{array}
\tag{81}
\end{equation}

If we now define the variable $Q=Q(x_{a}^{\mu };\tau )$ as

\begin{equation}
Q=\frac{1}{2}\varepsilon ^{ab}\delta _{\star }x_{a}^{\mu }x_{b}^{\nu }\eta
_{\mu \nu }-\delta \tau H_{T}-\delta _{\star }\Lambda ,  \tag{82}
\end{equation}%
then we find that (81) can be written as

\begin{equation}
\int_{\tau _{i}}^{\tau _{f}}d\tau \bigg \{\frac{d}{d\tau }{Q}+\varepsilon
^{ab}\dot{x}_{a}^{\mu }\delta _{\star }x_{b}^{\nu }\eta _{\mu \nu }+\delta
\tau \dot{H}_{T}-\delta _{\star }H_{T}\bigg\}=0  \tag{83}
\end{equation}%
or

\begin{equation}
\int_{\tau _{i}}^{\tau _{f}}d\tau \bigg \{\frac{d}{d\tau }{Q}+\varepsilon
^{ab}\dot{x}_{a}^{\mu }\delta x_{b}^{\nu }\eta _{\mu \nu }-\delta H_{T}%
\bigg\}=0.  \tag{84}
\end{equation}%
These expressions are of course the analogue of (11) or (13) respectively.
Thus, following a similar procedure as in section 2, we can prove that $Q$
plays a double role: a conserved quantity or a generator of canonical
transformations depending on whether the Hamilton equations of motion hold
or not.

Let us apply these results to two time physics. First we observe that in
terms of coordinates $x_{a}^{\mu }$ the definition (22) for the Poisson
brackets become%
\begin{equation}
\big\{f,g\big \}=\varepsilon _{ab}\frac{\partial f}{\partial x_{a}^{\mu }}%
\frac{\partial g}{\partial x_{b\mu }},  \tag{85}
\end{equation}%
for any canonical functions $f(x_{a}^{\mu })$ and $g(x_{a}^{\mu })$. Thus,
we find

\begin{equation}
\big\{x_{a}^{\mu },x_{b}^{\nu }\big \}=\varepsilon _{ab}\eta ^{\mu \nu }. 
\tag{86}
\end{equation}%
From this result is straightforward to check that the constraint $\Omega
_{ab}$ given in (68) gives

\begin{equation}
\big\{\Omega _{ab},\Omega _{cd}\big \}=C_{abcd}^{ef}\Omega _{ef},  \tag{87}
\end{equation}%
where the structure constants $C_{abcd}^{ef}$ are given by

\begin{equation}
\begin{array}{c}
C_{abcd}^{ef}=\frac{1}{2}\big[\varepsilon _{ac}(\delta _{b}^{e}\delta
_{d}^{f}+\delta _{d}^{e}\delta _{b}^{f})+\varepsilon _{ad}(\delta
_{b}^{e}\delta _{c}^{f}+\delta _{c}^{e}\delta _{b}^{f}) \\ 
\\ 
+\varepsilon _{bc}(\delta _{a}^{e}\delta _{d}^{f}+\delta _{d}^{e}\delta
_{a}^{f})+\varepsilon _{bd}(\delta _{a}^{e}\delta _{c}^{f}+\delta
_{c}^{e}\delta _{a}^{f})\big].%
\end{array}
\tag{88}
\end{equation}%
The expression (87) establishes that the constraint $\Omega _{ab}$ is in
fact a first class constraint.

Consider the variable

\begin{equation}
Q=\xi ^{ab}(\tau )\Omega _{ab},  \tag{89}
\end{equation}%
where $\xi ^{ab}=\xi ^{ba}$ are infinitesimal parameters. According to the
previous discussion this variable should be a conserved quantity or the
generator of gauge transformations depending on whether the equations of
motion are satisfied or not. In fact, using the formulae

\begin{equation}
\delta x_{a}^{\mu }=\big\{x_{a}^{\mu },Q\big \}  \tag{90}
\end{equation}%
and%
\begin{equation}
\delta _{\star }H_{T}-\frac{\partial }{\partial t}{Q}=0,  \tag{91}
\end{equation}%
which can be derived from (84) when the equations of motion are not
satisfied, we obtain that the constraint $\Omega _{ab}$ generates the
transformations

\begin{equation}
\delta x_{a}^{\mu }=\varepsilon _{ab}\xi ^{bc}x_{c}^{\mu }  \tag{92}
\end{equation}%
and

\begin{equation}
\delta \lambda ^{ab}=\dot{\xi}^{ab}-\xi ^{ef}\lambda ^{cd}C_{efcd}^{ab}. 
\tag{93}
\end{equation}%
We recognize in the expression (92) the infinitesimal transformation
associated with the group $Sp(2,R)\cong SL(2,R)$ with infinitesimal
parameter $\varsigma _{a}^{c}=\varepsilon _{ab}\xi ^{bc}$. Thus, we have
proved that if the Lagrange multipliers variation $\delta \lambda ^{ab}$ is
given by (93) then the action (67) is invariant under the $Sp(2,R)$ gauge
transformation (92). The remarkable fact is that this $Sp(2,R)$ invariance
of the action (67) is generated by the conserved quantity (89) corresponding
to the first class constraint $\Omega _{ab}$.

\bigskip\ 

\noindent \textbf{7.- Final remarks}

\smallskip\ 

In this work we revisited the Noether's first and second theorem. One of the
novel features of our presentation is that the canonical transformations can
be obtained directly from the action when the time derivative of quantity $Q$
is different from zero and the equations of motion are not satisfied. We
proved that our method may reveal hidden symmetries in specific cases.

As an application of our formalism we considered the cases of a relativistic
point particle and {the Friedberg \textit{et al.} model. }On the other hand,
since in two time physics the phase space has a unified character in the
sense that the spacetime and the momentum space are put together at the same
level, we found that an application of our formalism in this context
requires a generalization of the usual Noether's procedure. As a consequence
of such a generalization we show explicitly how the gauge transformations
for the coordinates and momenta, generated by the Hamiltonian constraint
associated with two time physics, also exhibit a unified character.
Moreover, using our method we have proved that the conserved quantity $Q$
given in (89) written in terms of the first class constraint $\Omega _{ab}$
generates the gauge symmetry of the action (67), clarifying further the
origin of the manifest gauge $Sp(2,R)$ symmetry.

The simplest possible action (67) corresponds to the "free" theory, in the
sense of the flat metric $\eta _{\mu \nu }$. Further generalization to an
interacting theory is possible by including gravitational background, gauge
fields, other potentials and higher-spin fields [19]. In those generalized
cases the constraint $\Omega _{ab}=x_{a}^{\mu }x_{b}^{\nu }\eta _{\mu \nu }$
is extended by including a curved metric $g_{\mu \nu }(x)$, gauge potential $%
A_{\mu }(x)$, a potential $U(x)$ and higher spins fields. In order to insure
the invariance $Sp(2,R)$ of the corresponding action the constraint (87) is
imposed as a differential constraint. An interesting aspect is that those
generalized cases of two time physics give all possible background fields in
one time physics (see Refs. [19] and [21] for details).

It may be interesting for further research to analyze the gauge fixing and
quantization of constrained Hamiltonian systems [28] from the point of the
present formalism. Another possible application of our formalism is related
to the connection between oriented matroid theory [29] (for application \ of
matroid theory on high energy physics see [30]-[32]) and two time physics.
In fact, via the chirotope concept in Refs. [25]-[27] were shown that there
is a deep connection between oriented matroid theory and two time physics.
Therefore it seems attractive to establish a connection between the concepts
of chirotope and gauge symmetry using the Noether theorem as discussed in
this work. Recently a complete analysis of Dirac's conjecture has been given
[33]. In this analysis the Legendre map from the tangent bundle and the
cotangent bundle (phase space) plays an essential role. In turn cotangent
bundle provides an example of geometrical structure of fiber bundles. It may
seems interesting to find a connection between such a geometrical structures
and the present formalism.

\bigskip

\noindent \textit{Note added}: After we finished this work, we received an
e-mail from I. Bars in which he called our attention about the footnote 3 on
Ref. [34]. In fact, in such a footnote there is also a discussion about the $%
Q$ conserved quantity. However, such a discussion is made at the level of
the Lagrangian and not on the full action. Moreover, $Q$ is assumed, from
the beginning, to be a generator of the canonical transformations.

\bigskip

\begin{center}
\textbf{Acknowledgments}
\end{center}

We would like to thank I. Bars, J. M. Pons, L. Lusanna, J. L. Lucio-Martinez
and H. Villegas for helpful comments. V. M. V. whishes to thank financial
support from Universidad Michoacana through Coordinaci\'{o}n de Investigaci%
\'{o}n Cient\'{\i}fica CIC 4.14 and CONACyT project 38293-E.

\end{document}